# A strange metal in a bosonic system

Chao Yang[1#], Haiwen Liu[2#], Yi Liu[3], Jiandong Wang[1], Sishuang Wang[1], Yang Wang[1], Qianmei He[1], Yue Tang[3], Jian Wang[3,4], X.C. Xie[3], James M. Valles Jr.[5*], Jie Xiong[1*] & Yanrong Li[1,6]

[1]State Key Laboratory of Electronic Thin Films and Integrated Devices, University of Electronic Science and Technology of China, Chengdu 610054, China.
[2]Center for Advanced Quantum Studies, Department of Physics, Beijing Normal University, Beijing 100875, China.
[3]International Center for Quantum Materials, School of Physics, Peking University, Beijing 100871, China.
[4]Beijing Academy of Quantum Information Sciences, Beijing 100193, China.
[5]Department of Physics, Brown University, 182 Hope Street, Providence, RI 02912, USA.
[6]Sichuan University, Chengdu 610065, China.
[#]These authors contributed equally to this work.
[*]Corresponding author. E-mail: jiexiong@uestc.edu.cn (J.X.); james_valles_jr@brown.edu (J.M.V.J).

## Abstract

**Fermi liquid theory forms the basis for our understanding of the majority of metals, which is manifested in the description of transport properties that the electrical resistivity goes as temperature squared in the limit of zero temperature. However, the observations of strange metal states in various quantum materials[1–21], notably high-temperature superconductors[1–10], bring this spectacularly successful theoretical framework into crisis. Distinct from the quadratic temperature dependence of the electron scattering rate ($1/\tau$) for ordinary metals, strange metals exhibit resistivity that scales linearly with temperature, indicating that the independent quasiparticle approximation in existing theoretical treatment is no longer valid. When $1/\tau$ hits its limit, $k_B T/\hbar$ where $\hbar$ is the reduced Planck's constant, $T$ represents absolute temperature and $k_B$ denotes Boltzmann's constant, Planckian dissipation[3,11,12,22–26] occurs and lends strange metals a surprising link to black holes[27], gravity[28–31], and quantum information theory[24]. While this strange metal phenomenology originates from investigations of only electronic phases, the centrality of a scattering rate dependent only on fundamental constants raises the question of whether it is exclusive to fermionic systems. Here, we show the characteristic signature of strange metallicity arising unprecedentedly in a bosonic system. Our nanopatterned YBa$_2$Cu$_3$O$_{7-\delta}$ (YBCO) film arrays reveal *T*-linear resistance as well as *B*-linear magnetoresistance over an extended temperature and magnetic field range in a quantum critical region in the phase diagram. Strikingly, the low-field magnetoresistance oscillates**

with a period dictated by the superconducting flux quantum of *h/2e* where *e* is the electron charge and *h* is the Planck constant, indicating that Cooper pairs instead of single electrons dominate the transport process and the system is bosonic. Moreover, the slope of the *T*-linear resistance $\alpha_{\text{cp}}$ appears bounded by $\alpha_{\text{cp}} \approx h/2e^2 \cdot 1/T_c^{\text{onset}}$ where $T_c^{\text{onset}}$ is the temperature at which Cooper pairs form, intimating a common scale-invariant transport mechanism corresponding to Planckian dissipation. In contrast to fermionic systems where the temperature and magnetic field dependent scattering rates combine in quadrature[15] of $\hbar/\tau \approx \sqrt{((k_B T)^2 + (\mu_B B)^2)}$, both terms linearly combine in the present bosonic system, i.e. $\hbar/\tau \approx (k_B T + \gamma \mu_B B)$, where $\gamma$ is a constant. By extending the reach of strange metal phenomenology to a bosonic system, our results suggest that there is a fundamental principle governing their transport which transcends particle statistics.

**Introduction**

The conventional theory of metallic transport predicts that the electrical resistivity $\rho(T)$ at low temperatures should vary as a function of $T^2$ due to the dominance of electron–electron scattering. However, this common understanding has encountered formidable conceptual challenges in accounting for some recent discoveries. One emerged soon after the discovery of high-temperature superconductors, in the form of a linear temperature dependence of their normal state resistivity over an exceptionally wide range of temperature. Termed by Anderson as one of the central dogmas for condensed matter physics[1], this unusual resistivity versus temperature structure and other anomalous properties signalled the breakdown of the quasiparticle picture in a region of the cuprate phase diagram now referred to as the strange metal phase[2–12]. Intriguingly, this phase is not confined to cuprates but is also present in many other strongly correlated electron materials, like heavy fermion metals[13,14,18–21], pnictides[15,17], twisted bilayer graphene[16], organic superconductors[17] and. Strikingly, near the quantum critical point (QCP), the resistance slope appears to be constrained by an upper bound $\alpha_F \approx h/2e^2 \cdot 1/T_F$ where $T_F$ is the Fermi temperature[1–15,17,18]. This bound corresponds to a scattering rate $1/\tau = k_B T/\hbar$, the so-called Planckian dissipation limit[11,12,22–26], which has been received considerable attention recently due to its implications for black holes, gravity, and quantum computation[27–31].

Another challenging discovery is the anomalous metallic state in quasi-2D superconducting materials that appears over a range of magnetic field or normal state resistance[32–43]. Characterized by saturated residual resistivity near zero temperature, the anomalous metallicity observed in different systems may provide insights necessary for extending our understanding of metals. The fact that Cooper pairs of electrons instead of single electrons dominate the transport in these different anomalous metals leads us to speculate that strange metals might be not restricted to fermionic systems, either.

Here, in a break with the existing pattern that strange metal states are exclusive to systems of fermions, we show that this unusual metallic behavior emerges in a bosonic host and thus the behavior transcends particle statistics. The observed *T*-linear resistance in a series of nanopatterned YBCO thin films involves clear bosonic transport as evidenced by the period of oscillations in their magnetoresistance. Near the quantum critical point of their superconductor-anomalous metal transition, the slope of the *T*-linear resistance $\alpha_{\text{cp}}$ of the current YBCO strange metal reaches a maximum $\alpha_{\text{cp}} \approx h/2e^2 \cdot 1/T_c^{\text{onset}}$. This apparent bound resembles the one in fermionic systems $\alpha_{\text{F}} \approx h/2e^2 \cdot 1/T_F$ which corresponds to the Planckian dissipation limit and defines the strange metal in their phase diagram. Moreover, the magnetoresistance is linear in this regime with a slope that appears bounded by $\beta_{\text{cp}} \approx h/2e^2 \cdot 1/B^*$ ($B^*$ is a characteristic magnetic field), intimating that a magnetic field introduces an independent and distinct scattering process that is not apparent in fermionic strange metals. Interestingly, the ratio of the slopes $\alpha_{\text{cp}}/\beta_{\text{cp}}$ appears only determined by fundamental constants $2\mu_B/k_B$, indicating scale-invariant behaviors. Our study represents a game-changing step in broadly advancing the conceptual understanding of the mysterious physics of the strange metals.

## Results

### *T*-linear resistance

The present studies were conducted on a series of high-*T*c superconducting YBCO films nanopatterned with a triangular array of holes. A series of these films with a range of normal state resistances $R_N$, measured just above the superconducting transition temperature, were produced using reactive ion etching (RIE) technique[42]. Nanoporous masks were placed atop pristine 12-nm-thick films and subjected to different etching doses to yield a triangular array of pores that are 103 nm apart and 70 nm in diameter (Supplementary Fig. 1). Increasing etching dose increased $R_N$ by inducing more side wall damage in the pores to make the links between array nodes more resistive.

As shown by the sheet resistance as a function of temperature measurements $R_S(T)$ in Fig. 1a, these arrays undergo a superconductor-anomalous metal-insulator transition with increasing $R_N$. The transport behavior that we focus on occurs in films near the anomalous metal-insulator phase boundary where $R_N$ is close to $R_Q$. For $R_N$ less than ~3kΩ, the $R_S(T)$ exhibit relatively sharp transitions to a superconducting state starting at an onset temperature $T_c^{onset}$, that is close to the pristine film transition temperature. As $R_N$ grows larger than ~3kΩ, the anomalous metal phase for which $R_S(T)$ saturates at low temperatures to a constant value appears[42] (Supplementary Fig. 2). When $R_N$ approaches 12kΩ, near the anomalous metal-insulator transition point, the $R_S(T)$ develops a nearly linear temperature dependence over an extended range, as shown by five representative samples, f1–f5. Strikingly, in the case of f3 ($R_N$ approaches 12kΩ) the linear temperature dependence extends over a factor for 20 in temperature from 60 K to 3 K with a low-temperature intercept near zero. Above 12 kΩ, the linear temperature dependence develops into an insulating dependence (i.e. $dR/dT < 0$) in the low-temperature limit.

Having detected a regime with *T*-linear resistance, we now turn to investigate its slope. Denoting it as $\alpha_{cp}= dR_S(T)/dT$, we introduce a normalized form $\eta = (2e^2/h) \cdot \alpha_{cp} T_c^{onset}$, inspired by $\eta = 2e^2/h \cdot \alpha_F T_F$ in fermionic systems[7]. The extracted $\alpha_{cp}$ and $\eta$ values for each sample measured in Fig. 1a are plotted against $R_N$ in Fig. 1b. The slope peaks at a value of $\alpha_{cp} \approx 157 \, \Omega \, K^{-1}$ and $\alpha$ = 1.02 for the film f3 exhibiting the greatest range of *T*-linear resistance. Over the $R_N$ range bracketing f3, 10 kΩ to 15 kΩ, the slope remains within 25% of the maximum value while the lowest temperature behaviors of the $R_S(T)$ evolve from anomalous metal (e.g. f1 and f2) to insulator (e.g. f5).

**Bosonic nature**

Information about the charge carriers producing the linear resistance comes from the low magnetic field magneto-resistance. Below $T_c^{onset}$, the magnetoresistances of the nanopatterned thin films (Fig. 1c and Supplementary Fig. 3) oscillate with a constant period of $\mu_0 H_{period} \approx$ 0.225T, corresponding to one superconducting flux quantum per unit cell of the array with $H_{period} \approx h/2eS$ and the unit cell area $S \approx$ 9200 nm². These quantum oscillations below $T_c^{onset}$ provide direct evidence that Cooper pairs instead of single electrons dominate the transport process[42,44].

## *B*-linear resistance

The *T*-linear resistance persists in magnetic fields (*B* > 1 T) beyond the magnetoresistance oscillations, as shown in Fig. 2a–c. The magnetoresistance is uniformly positive up to the measurement limit of 9T. Examining the magneto-resistance directly reveals that it is nearly linear over this field scale for a wide range of temperatures (Fig. 2d, e and Supplementary Fig. 4). By contrast, the magneto-resistance for the lower $R_N$ superconducting (SC) sample is non-linear (Fig. 2f). Notably, the linear magneto-resistance curves of samples f1–f5 are approximately parallel for a wide range of temperatures. Their slopes obtained from linear fits, $\beta_{cp} = \frac{dR}{dB}$, as a function of temperature capture this observation (Fig. 2g, h and Supplementary Fig. 5). $\beta_{cp}$ for film f4, for example, shows a plateau over which it varies by only 12% from 30K to 2K, while its resistance changes by almost a factor of 3 as shown in Fig. 2g. The temperature range for this plateau or broad peak coincides with the temperature range over which the resistance is linearly temperature dependent. Writing the magnetoresistance in a form similar to the temperature dependence $\frac{dR}{dB} = \frac{h}{2e^2} \cdot \frac{1}{B^*}$, we can identify a magnetic field scale $B^*$ for the slope. At the maximum slope, $B^*$ is approximately 60 T.

The linear dependencies of the sheet resistance on *T* and *B* indicates that：

$$(R_S(T,B) - R_S(0,0))/T = \alpha_{cp} + \beta_{cp} \cdot \frac{B}{T} \quad (1)$$

over a wide temperature and magnetic field range. Replotting the $R_S(T,B)$ as $(R_S(T,B) - R_S(0,0))/T$ vs. $\frac{B}{T}$ leads to the collapse onto a single curve from 3 K to 30 K as shown for f4 in Fig. 3a. The scaling plot of other samples are shown in Fig. 3b and Supplementary Fig. 6. The ratio of $\beta_{cp}$ and $\alpha_{cp}$ gives a quantity that is very close to a ratio of fundamental constants $k_B/\gamma\mu_B$, where $\gamma$ is a dimensionless parameter, and $\mu_B$ is Bohr magneton.

## Phase diagram

To depict the region over which the scaling appears to hold, we plot $1/R_N \cdot d^2R_S/dT^2$ on a color scale on $R_N$ versus *T* axes, as shown in Fig. 4a. This region is represented by the wide green region in the central part Fig. 4a, where $1/R_N \cdot d^2R_S/dT^2$ is near zero. It funnels down with decreasing temperature from $T_c^{onset}$ becoming narrow near the critical sheet resistance $R_{Nc} \sim h/2e^2$ separating the insulating and anomalous metal phases at low temperatures. We label this green area as the bosonic strange metal, reminiscent of the strange metal in fermionic systems.

**Discussion**

The present work uncovers a distinctive bosonic metallic transport behavior that deviates from the previous boson localization paradigm of the superconductor-insulator transition in which metallic behavior appears with a critical resistance $R_Q$ at a critical point of a tuning parameter like magnetic field or normal state resistance. Like the anomalous metal state, this strange metal state exists over a range of tuning parameters with residual resistances far below $R_Q$. The primary characteristics that distinguish this bosonic transport behavior are 1) a resistance that changes linearly with temperature with a slope that appears bounded by a quantum of resistance and is inversely proportional to an energy scale characterizing the ground state of the charge carriers, 2) a magnetoresistance that changes linearly with magnetic field as implied by the data collapse in Fig. 3, and 3) the appearance of these behaviors in a fan-shaped region that appears to terminate in a quantum critical point in the phase diagram in Fig. 4.

Because these bosonic transport characteristics are similar to the fermionic strange metals[1–19], we discuss them in that context first. In the model based on the Drude approach, the scattering rate is presumed to reach and be bounded by bounded by the so-called Planckian dissipation limit, $1/\tau = k_B T/\hbar$. The slope of the sheet resistance, $\alpha = (m^*/ne^2)(k_B/\hbar)$ then only depends on $m^*/n$ where $m^*$ is the effective mass and $n$ is the carrier density. For the fermionic systems, $n/m^*$ is proportional to $T_F$ in a two-dimensional approximation. For Cooper pairs as the carriers, $T_c^{onset}$, which measures the portion of the Fermi volume that forms pairs serves as the appropriate energy scale. Thus, the linear temperature dependence of the bosonic system is much larger than the fermionic systems by the ratio $T_F/T_c^{onset}$, in accord with the data (also see Table 1). Turning to the magnetic field dependence, fermionic systems show $\hbar/\tau \approx \sqrt{((k_B T)^2 + (\mu_B B)^2)}$ in some cases[15]. This scale free form indicates that the scattering does not depend directly on any intrinsic energy scale characteristic of the systems. The data for the bosonic system here are also consistent with a scale free scattering rate with $\hbar/\tau \approx k_B T + \gamma \mu_B B$ with $\gamma \approx 2$. This form, in contrast to the quadrature form for the fermionic case, implies that there are two independent scattering processes.

Reminiscent of Planckian dissipation $\hbar/\tau_1 \approx k_B T$, it implies a magnetic field damping rate $\hbar/\tau_2 \approx \gamma \mu_B B$, namely, the type-II Planckian dissipation. This conjecture relies on the experimental observation that $\gamma = \frac{k_B T_c^{onset}}{\mu_B B^*}$, which is remarkably close to the Landau *g* factor.

Viewing this novel behavior within the context of the natural model for these arrays, the quantum XY model with dissipation provides an explanation for why this novel behavior has not been observed before. The dissipative processes involve the coupling of the bosons to low-lying fermionic excitations of this d-wave system[42,45]. This dissipative model can exhibit critical behavior[46] that differs from the non-dissipative models[47]. It has three, rather than two, ground states: an ordered superconducting state, a disordered insulating state and in between a quasi-ordered state that is of interest here[48,49]. The phase of the order parameter in the less familiar quasi-ordered state is spatially ordered and temporally disordered[48,49]. Here, we tune the normal resistance $R_N$ to cross the three phases. Estimates of the model parameters based on the geometry of these arrays and a standard form for the Josephson coupling energy between array nodes leads to the identification of the phases at different $R_N$ that is shown in Fig. 4b. The absence of reports of strange metal behavior in prior superconducting array experiments suggests that its emergence here is related to the high-*T*c superconducting state. Previous investigations employed primarily s-wave superconductors, which afford fewer of the low energy excitations for the dissipation necessary to enter the critical region described above. Also, their anomalous metal behaviors appear at significantly lower $R_N$ where the BKT transition appears closer to $T_c^{onset}$[32]. While this explanation is appealing, it could be that the higher $T_c^{onset}$ made the linear-*T* resistance more apparent as it extends over a decade in temperature.

Future measurements of the frequency dependent conductivity show promise for direct insights into the dynamics of this system that could reveal the origins of these linear transport phenomena. A measurement of the Drude width can be particularly interesting as it has been proposed to scale with the entropy in models proposed to apply to matter in many forms including cold atoms, the solid-state materials and quark gluon plasmas. In addition, as dissipation induced decoherence processes are important in quantum computing, our results provide a platform for further study of the dissipative quantum system.

**Methods**

## Sample preparation

The high-$T_c$ YBa$_2$Cu$_3$O$_{7-\delta}$ superconducting films were epitaxial grown on (001)-oriented substrates SrTiO$_3$ in a dc magnetron sputtering system. The films were grown in an argon/oxygen (ratio 2:1) mixture under the working pressure of 30 Pa at $T \approx 700°C$, with a very low growth rate (~0.8 nm min$^{-1}$). Then, an anodic aluminum oxide (AAO) template of 200 nm thick was transferred to YBCO thin film. By reactive ion etching, the AAO pattern of a triangular array of holes with a 70-nm diameter and 103-nm period is duplicated onto the YBCO films. The etching recipe includes 30 sccm CHF$_3$, 40 sccm Ar, 20 mTorr (1 Torr ≈ 133 Pa) pressure, 200 W rf power, 1800W inductive coupling plasma (ICP). The YBCO films can be tuned from superconducting state to an insulating state by increasing the etching time from 20s to 160s[42].

## Measurements

The in-plane resistance and magnetoresistance were measured with a standard four-probe in a commercial Physical Property Measurement System (PPMS, Quantum Design) with magnetic fields up to 9 T. The size of YBCO samples for the measurements is typically 1 cm × 1 cm. For the transport measurements, the excitation electrical current as low as possible (5 nA~500 nA) was applied in the ab-plane while the magnetic field was applied along the c-axis. The sweeping rate is typically 0.04 T min$^{-1}$ for measuring the magnetoresistance quantum oscillations.

## Acknowledgements

The authors would like to thank C. M. Varma, H. Yao, A. Lucas, and J. M. Kosterlitz for the fruitful discussions. This work was supported by the National Natural Science Foundation of China (grant 51722204, 11888101, 50902017, U20A20244, 12022407, and 11774008), the National Basic Research Program of China (grant 2018YFA0305604, 2017YFA0303300, and 2017YFA0304600), Beijing Natural Science Foundation (Z180010).

## Author contributions

J.X. and J.M.V.J. conceived the study and supervised the project together with Y.L. and J.W.. C.Y. and JIAND.W. fabricated the samples. C.Y., Y.LIU., S.W., Y.W., Q.H. and Y.T. performed the experimental measurements. C.Y., H.L., J.M.V.J. and J.X. analyzed the data with the contribution of Y.LIU., J.W. and Y.L.. X.C.X. participated in discussions. C.Y., J.M.V.J., H.L, J.X. and Y.L wrote the manuscript with comments from J.W..

## Competing interests

The authors declare no competing interests.

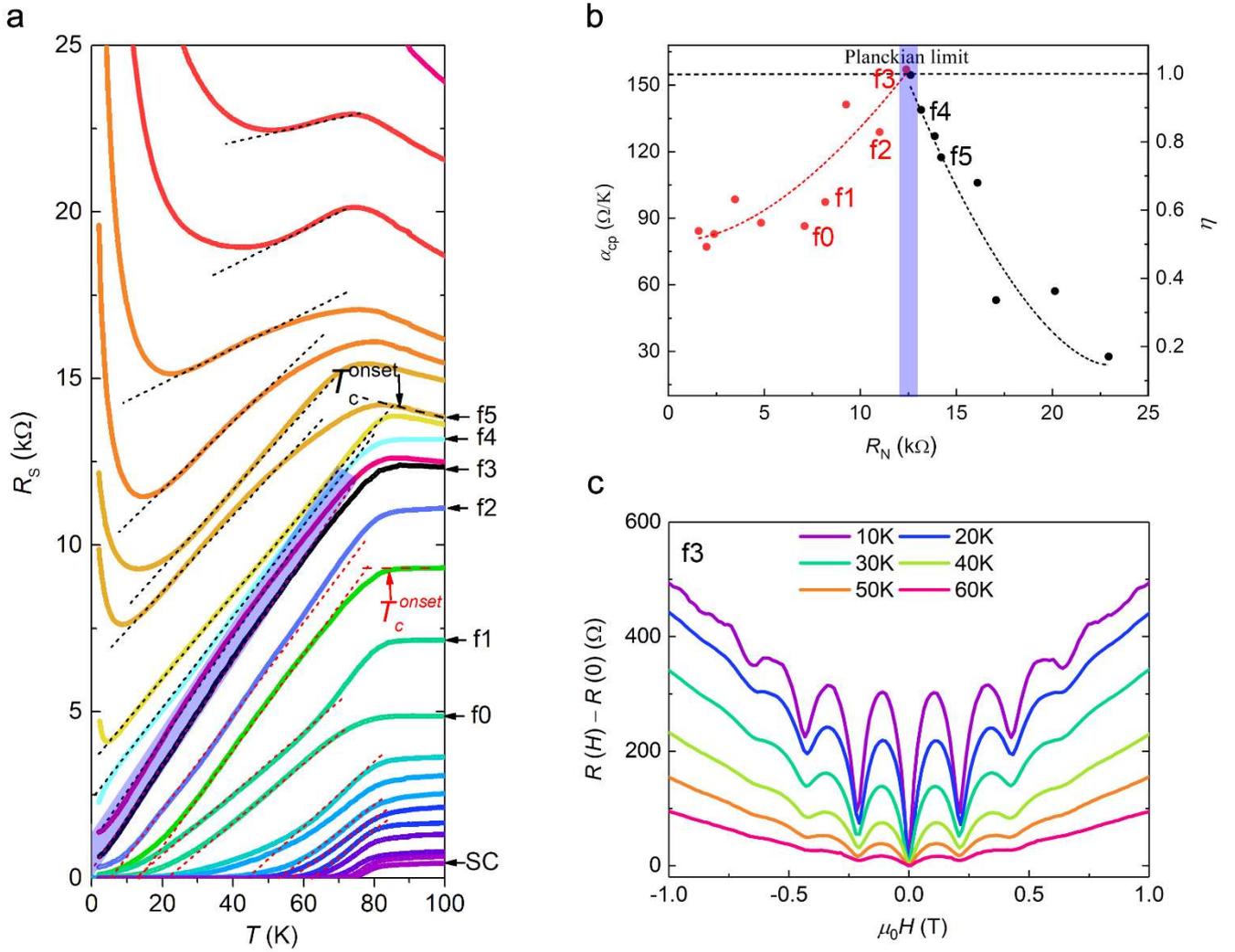

**Fig. 1| Linear-in-temperature resistance near a bosonic anomalous metal-insulator transition in nanopatterned YBCO thin films.** (**a**) Sheet resistance as a function of temperature $R_S(T)$ for a series of nanopatterned YBCO films over a range of normal state sheet resistances $R_N$. The dashed lines are linear fits to the data to determine the slope $\alpha_{cp}$ of the linear portion of the $R_S(T)$ that develops for $R_N$ close to $R_Q$. Close to the transition point of anomalous metal-insulator transition, the resistance shows linear temperature dependence from around 60 K down to around 3 K. (**b**) Slopes $\alpha_{cp}$ in a) as a function of $R_N$. The right axis $\eta$ is the slope normalized by $(2e^2/h) \cdot T_c^{onset}$, where $T_c^{onset}$ is defined by the construction shown. The maximum observed slope, deemed the Planckian limit, corresponds to $\eta=1$. (**c**) Low field magnetoresistance of film f3 ($R_N \sim h/2e^2$) as a function of temperature. The damped oscillations have a period $\mu_0 H_0 \sim 0.225T$ corresponding to one charge 2e flux quantum per unit cell of the array. (other representative films are shown in Supplementary Fig. 3).

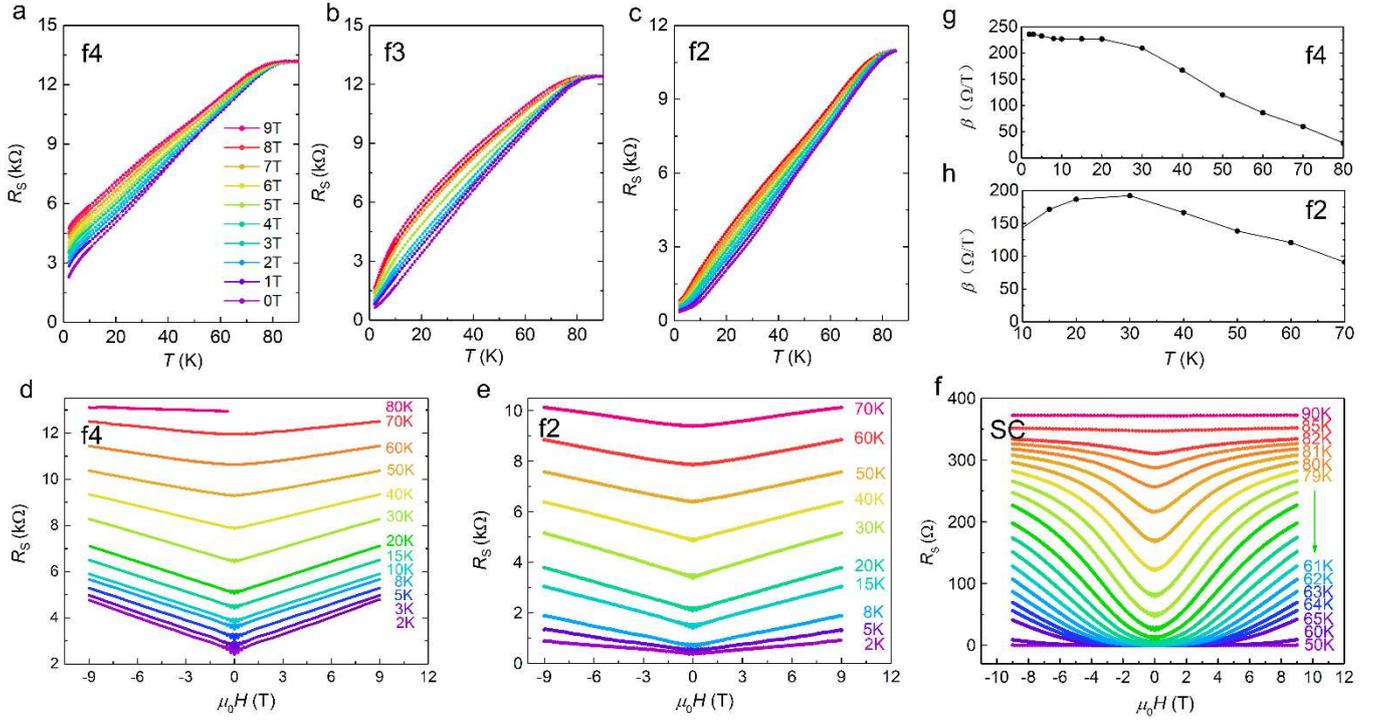

**Fig. 2| *T*-linear resistance and scale-invariant *B*-linear resistance in nanopatterned YBCO thin films under perpendicular magnetic field.** (**a–c**) $R_s(T)$ curves for films: f4 (**a**), f3 (**b**), f2 (**c**), under different perpendicular magnetic fields indicated by the legend in (a). (**d–f**) The magnetoresistance for films: f4 (**d**), f2 (**e**), SC (**f**). The slope of magnetoresistance $\beta$ of f4 (**a**) and f2 (**d**) as a function of temperature.

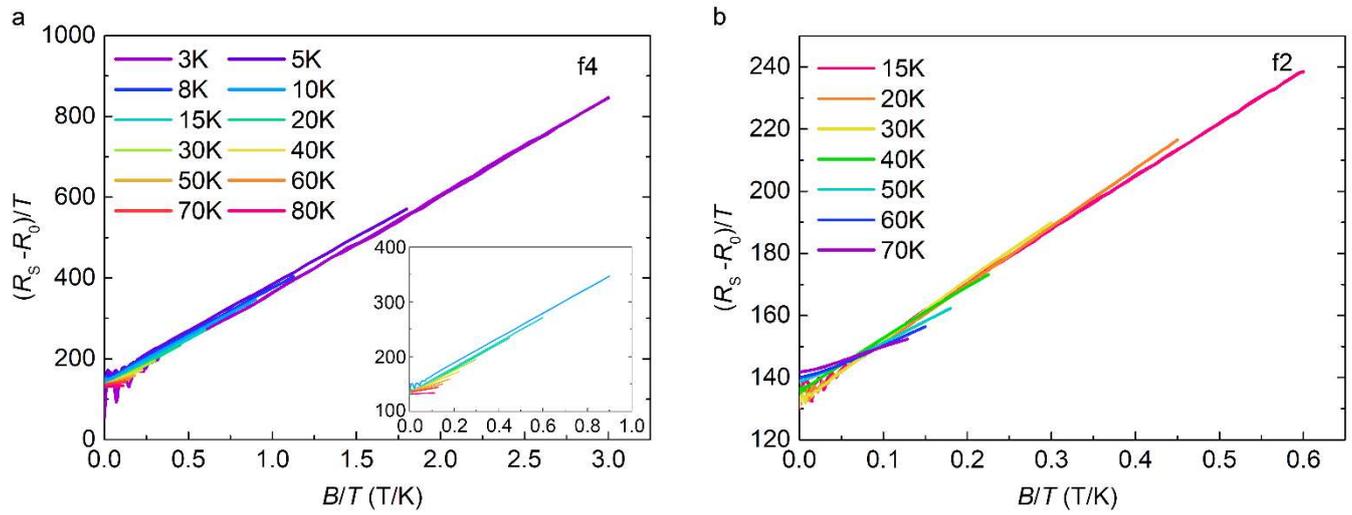

**Fig. 3| *B-T* scaling in nanopatterned YBCO thin films.** The resistance of f4 (**a**) and f2 (**b**) after subtracting the residual resistance $R_S(0,0)$, the reminder is divided by temperature, and plotted versus *B/T*. The insert is the detail of figure 3a from 10K to 80K. (The similar plots of other samples are shown in Supplementary Fig. 7).

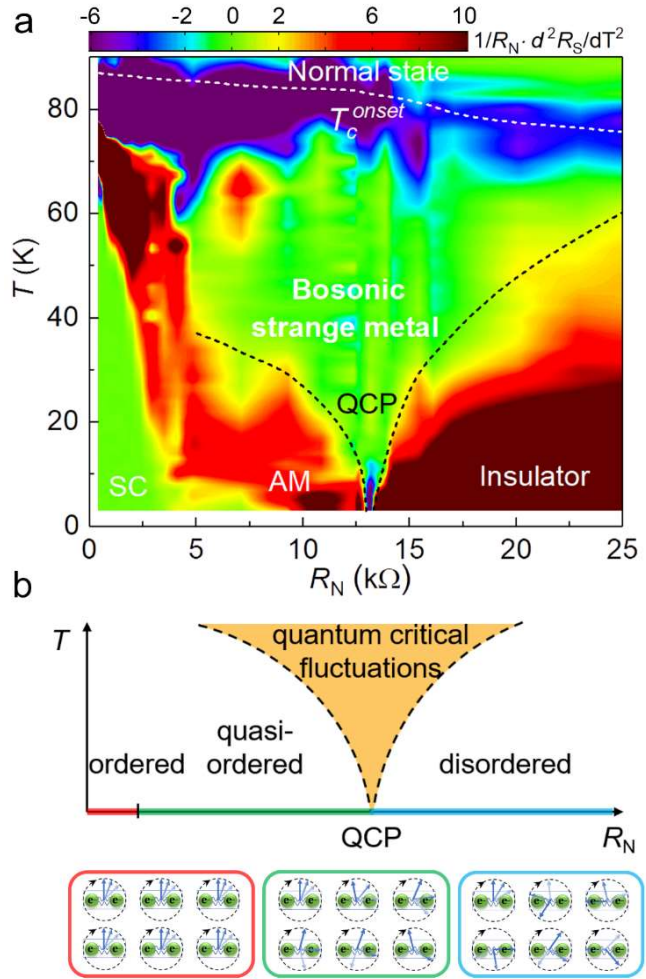

**Fig. 4| Phase diagram of nanopatterned YBCO thin films.** (**a**) The normalized second derivative of resistance $1/R_N \cdot d^2 R_S/dT^2$ versus temperature and the normal state resistance $R_N$. The normal state and other states are separated by $T_c^{onset}$. The color scale conveys the second derivative of the sheet resistance as a function of temperature normalized to $R_N$. The central green region ($1/R_N \cdot d^2 R_S/dT^2$ close to zero), labelled bosonic strange metal, corresponds to the regime where the temperature dependence of the resistance is linear. Its boundaries (black dashed lines) are drawn where the deviations from linearity become significant. The bordering regions evolve into the bosonic insulator on the right and the anomalous metal phase (AM) in the low temperature limit. The superconducting phase (SC) is the green region to the right where $1/R_N \cdot d^2 R_S/dT^2$ approaches zero because $R_S$ becomes very small. (**b**) Proposed idealized phase diagram illustrating a quantum critical region that corresponds to where strange metal behavior is observed. The bounding phases are the anomalous metal and bosonic insulator. The blue arrow stands for the SC order parameter. The ordered state (left), which is ordered both in space and time, corresponds to SC state. The quasi-ordered phase (middle), which is spatially ordered but disordered in time, corresponds to AM state. The disordered phase (right), which is disordered both in space and time, corresponds to insulator state. Near the critical point of quasi-ordered to disordered state, there is a quantum critical region where there is exotic *T*-linear resistance, corresponding to the bosonic strange metal.

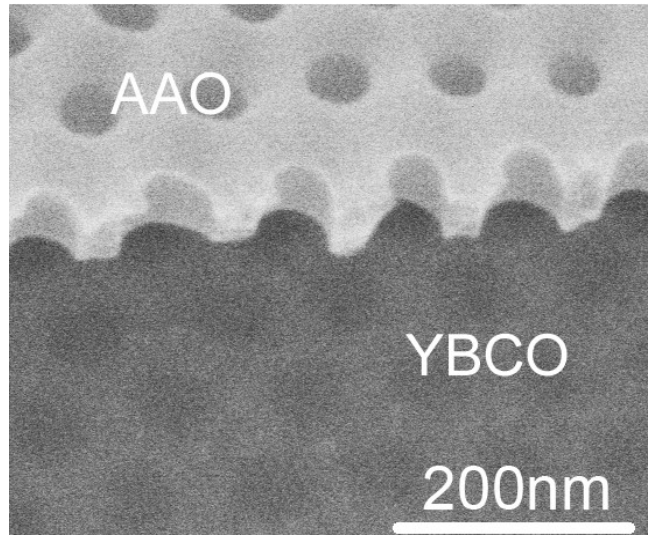

**Supplementary Fig. 1|** The scanning electron microscopy of a nanopatterned YBa$_2$Cu$_3$O$_{7-\delta}$ (YBCO) thin film. The 12-nm-thick nanopatterned YBCO thin film was fabricated by reactive ion etching through an anodic aluminum oxide (AAO) membrane directly placed atop the YBCO. By RIE, the anodized aluminum oxide pattern of a triangular array of holes with ~70-nm diameter and ~103-nm period was duplicated onto the YBCO film.

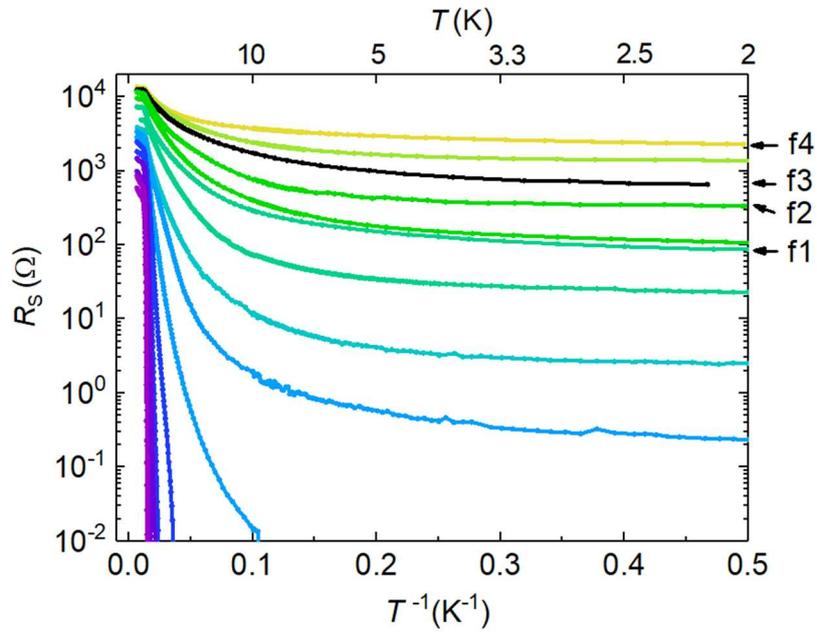

**Supplementary Fig. 2|** Arrhenius plot of the $R_S(T)$ data of anomalous metallic state and superconducting state in Fig. 1a. The resistance of the four representative anomalous metallic state films f1, f2, f3 and f4 that have a linear temperature dependence over an extended range saturates at low temperatures on a $1/T$ scale. This saturation is a characteristic of the intermediate anomalous metallic state in the superconductor-insulator transition nanopatterned YBCO thin films.

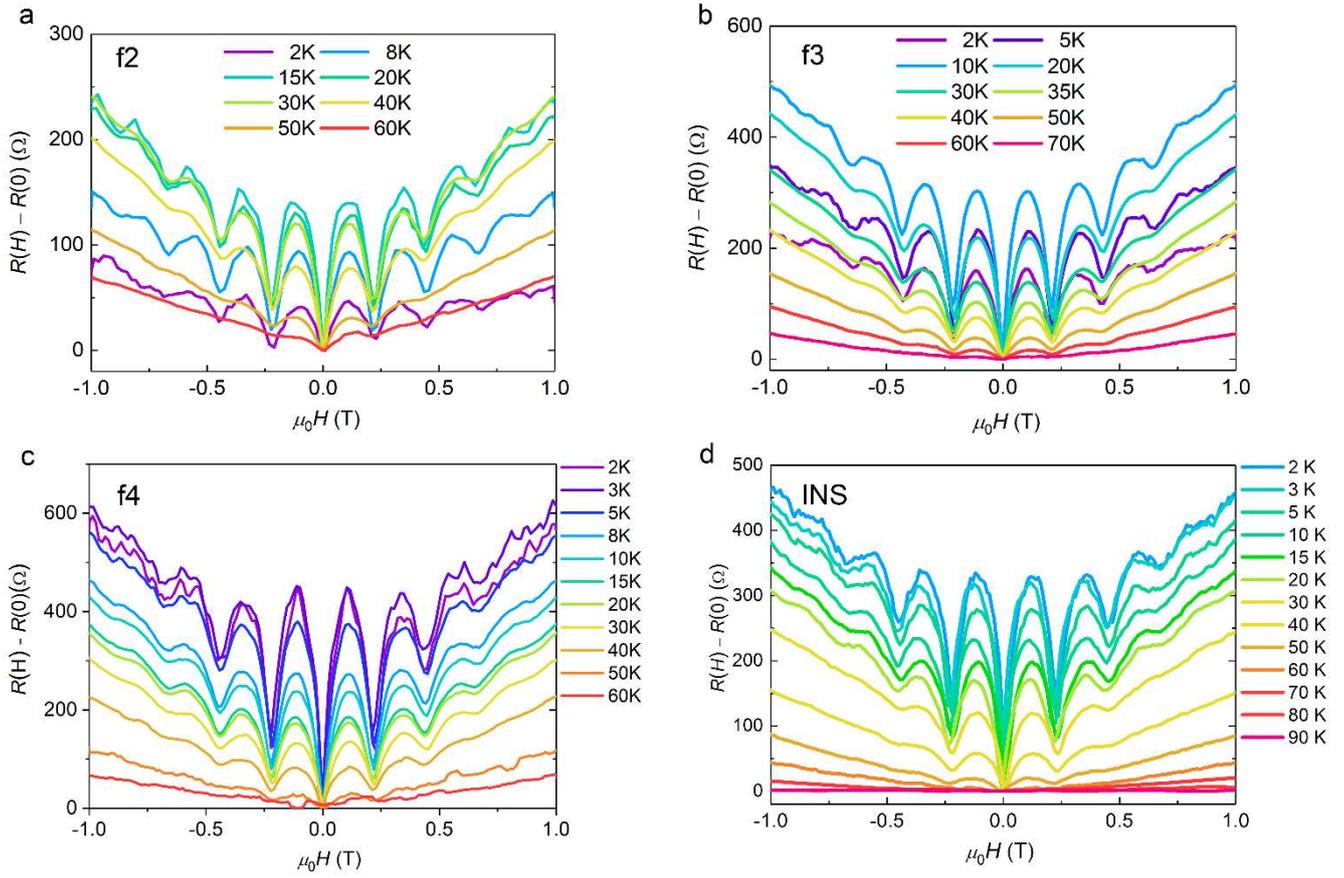

**Supplementary Fig. 3 | The magneto-resistance oscillations in nanopatterned YBCO thin film.** The full data of the magneto-resistance four representative YBCO thin films at various temperatures from 2K to 80K (**a**) f2 (**b**) f3 (**c**) f4 (**d**) a representative insulating state YBCO thin film. The observed oscillation period $\mu_0 H_0 \sim 0.225T$ is close to one superconducting flux per area of a unit cell that is $h/2eS$, where $h$ is Planck's constant, $e$ is the electron charge and $S$ is ~9200 nm². These h/2e magneto-resistance oscillations indicate that Cooper pairs participate in the transport below $T_c^{onset} \approx 84K$.

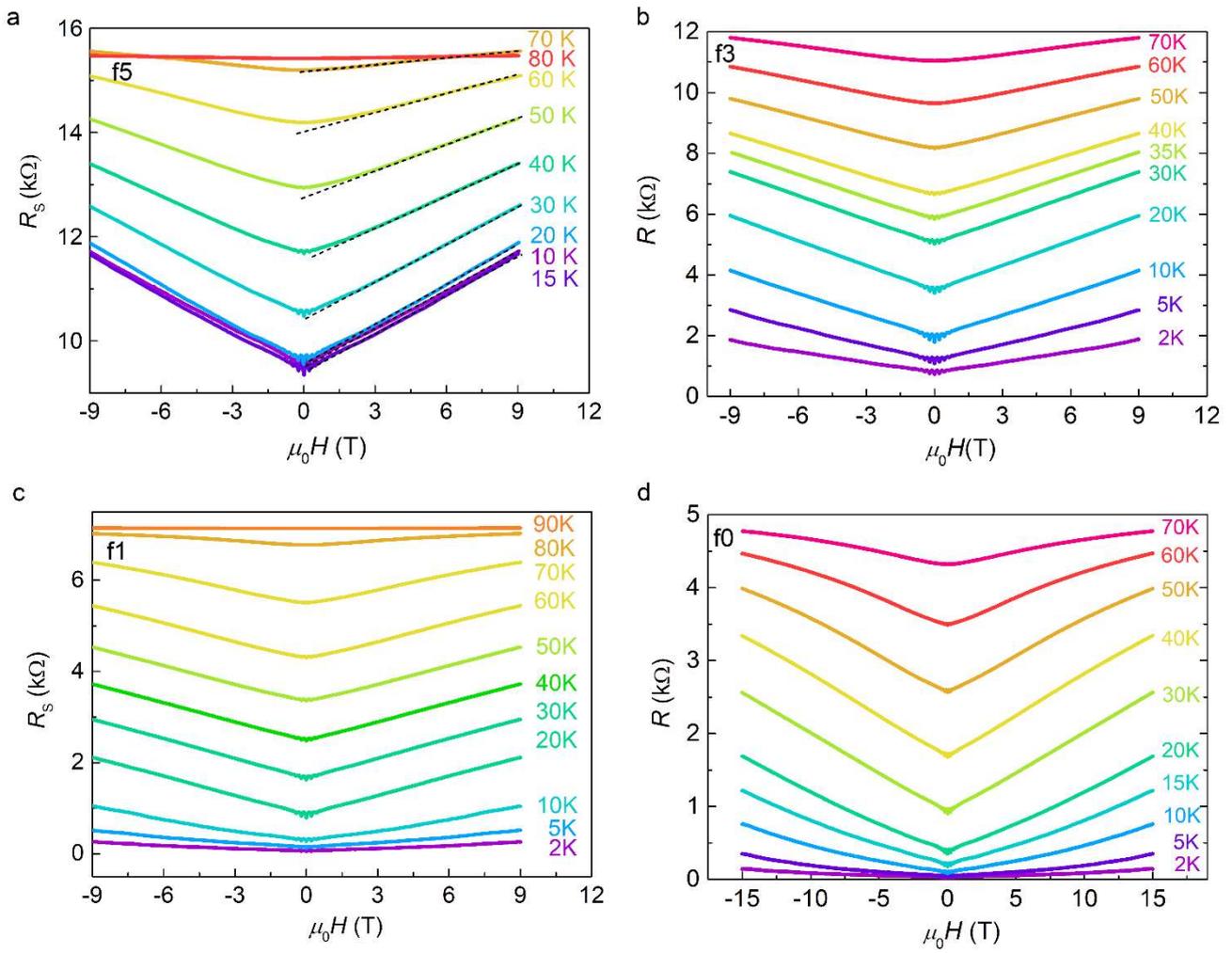

**Supplementary Fig. 4| scale-invariant *B*-linear resistance in nanopatterned YBCO thin films under perpendicular magnetic field. (a–d)** The magnetoresistance for films: f5 **(a)**, f3 **(b)**, f1 **(c)**, f0 **(d)**.

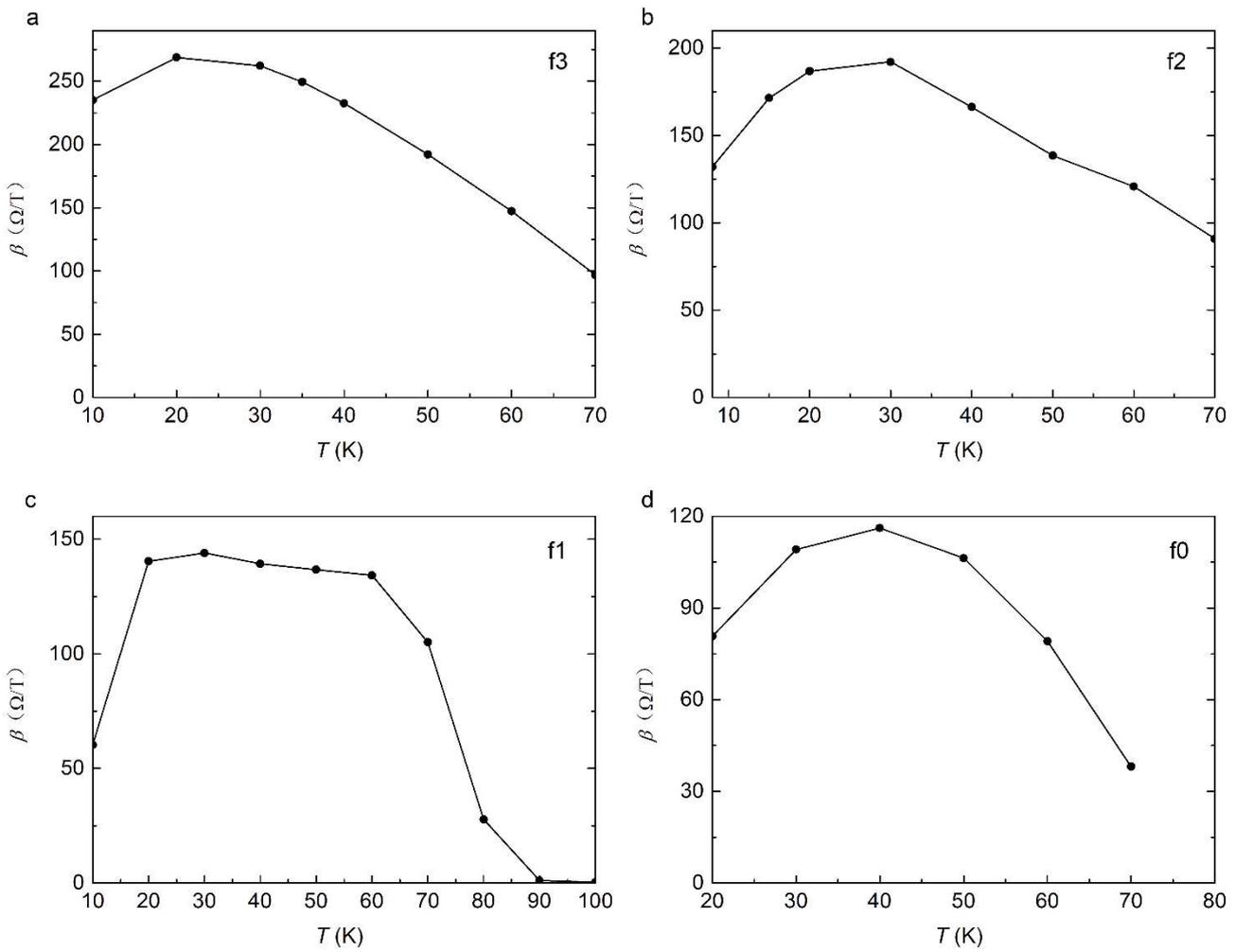

**Supplementary Fig. 5|** The slope of magnetoresistance $\beta$ as a function of temperature: f3 **(a)**, f2 **(b)**, f1 **(c)** and f0 **(d)**.

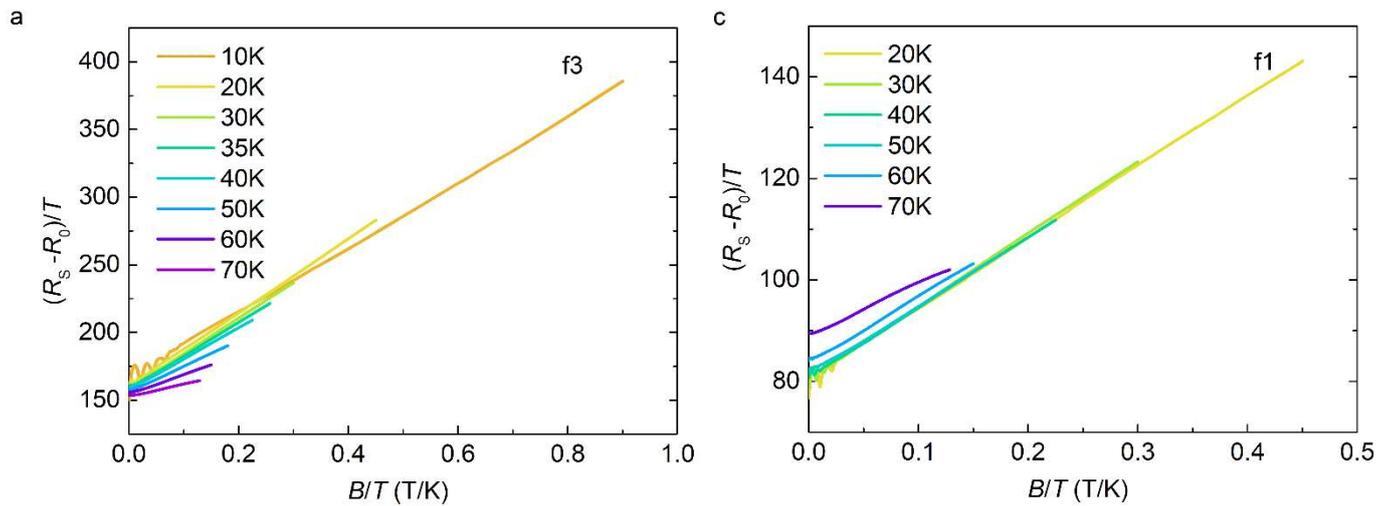

**Supplementary Fig. 6|** *B-T* scaling in nanopatterned YBCO thin films of f3 (**a**) and f1 (**b**).

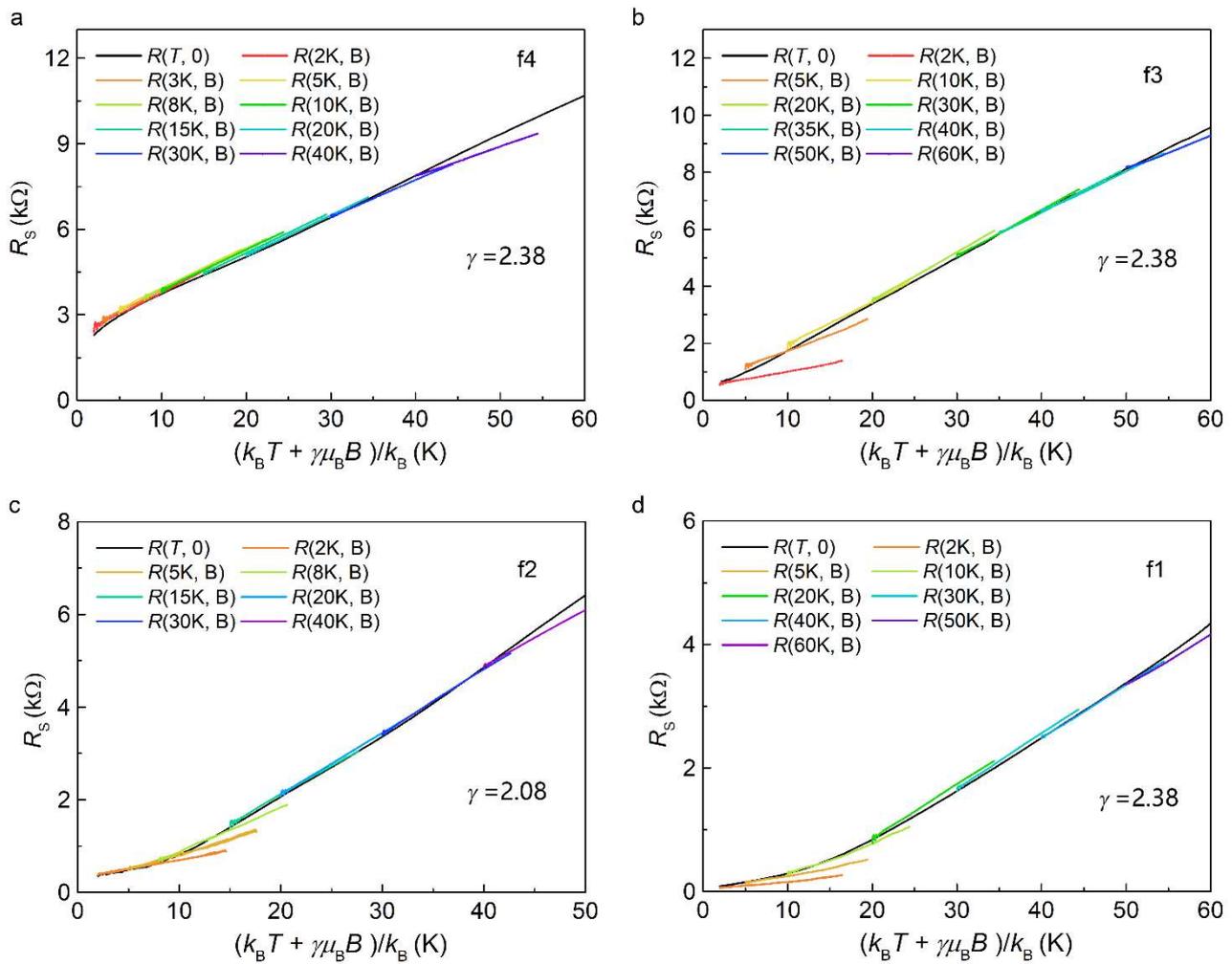

**Supplementary Fig. 7|** The resistance and magneto-resistance of f4 (**a**), f3 (**b**), f2 (**c**) and f1 (**d**) as a function of $(k_B T + \gamma \mu_B B)/k_B$, where $\gamma$ parameter can be estimated by adjusting it when the curves collapse best.

| | LSCO at p= 0.21 | Nanopatterned YBCO thin films near the anomalous metal to insulator transition |
|---|---|---|
| $T$-linear resistivity $R_S = \alpha T + R_0$ | √ | √ |
| Carriers | (Electrons) Fermions | (Cooper pairs) Bosons |
| $\alpha\ (\Omega \cdot K^{-1})$ | 15 | 157 |
| A renormalized value $\eta$ | $\eta = (2e^2/h) \cdot \alpha_F / T_F \approx 1$, where $\hbar/\tau = \eta k_B T$ | $\eta = (2e^2/h) \cdot \alpha_{cp} / T_c^{onset} \approx 1$ |

Extended Data Table 1 | Comparison between LSCO at p=0.21 and nanopatterned YBCO thin films near the anomalous metal to insulator transition.